\documentclass[lettersize,journal]{IEEEtran}
\IEEEoverridecommandlockouts
\usepackage{amsmath,amssymb,amsfonts}
\usepackage{algorithmic}
\usepackage{graphicx}
\usepackage{textcomp}
\usepackage{xcolor}

\usepackage{verbatim}
\usepackage[style=ieee,
bibstyle=ieee,
backend=biber,
minnames=1,
maxcitenames=2,
maxbibnames=6,
doi=true,
isbn=false,
url=false,
date=year,
]{biblatex} 

\usepackage{listings}

\usepackage{tcolorbox}
\usepackage{enumitem}
\usepackage{xcolor}
\usepackage{mdframed}
\AtEveryBibitem{\clearlist{language}}
\AtEveryBibitem{\clearlist{month}}
\usepackage{orcidlink}

\usepackage{cleveref}
\usepackage{float}
\def\BibTeX{{\rm B\kern-.05em{\sc i\kern-.025em b}\kern-.08em
    T\kern-.1667em\lower.7ex\hbox{E}\kern-.125emX}}
\begin{document}

\title{Automated vehicles: Challenging transition from Minimal Risk Condition}

\author{Wolfgang Kröger\orcidlink{0009-0004-6681-9235} and Leon Johann Brettin\orcidlink{0009-0007-2796-0860}
\thanks{
Wolfgang Kröger is with the Federal Institute of Technology Zurich (ETH), Zurich, Switzerland; and with the Swiss Academy of Engineering Sciences (SATW), Zurich, Switzerland (E-mail: \url{wkroeger@ethz.ch})

Leon Johann Brettin is with the Institute of Control Engineering,
Technische Universität Braunschweig, Braunschweig, Germany (E-Mail: \url{l.brettin@tu-braunschweig.de})

Both authors contributed equally to this article.
Corresponding author is Leon Johann Brettin}
}


\twocolumn[
  \begin{@twocolumnfalse}
    \begin{center}{\LARGE\bfseries IEEE Preprint Notice\par}\end{center}
    \vspace{2ex}
    \noindent This work has been submitted to the IEEE for possible
    publication. Copyright may be transferred without notice, after which this
    version may no longer be accessible.\par
    \vspace{2ex}
    \noindent Submitted to \emph{IEEE Transactions on Intelligent
    Vehicles}.\par
    \vspace{2ex}
    \noindent\textbf{Cite as:}\par
    \vspace{1ex}
    \noindent\fbox{\parbox{\dimexpr\linewidth-2\fboxsep-2\fboxrule\relax}{%
      W.~Kr\"oger and L.~J.~Brettin, ``Automated vehicles: Challenging
      transition from Minimal Risk Condition,'' \emph{IEEE Transactions on
      Intelligent Vehicles}, 2026, submitted for publication.}}\par
    \vspace{2ex}
    \noindent\textbf{BibTeX:}\par
    \vspace{1ex}
\begin{lstlisting}
@article{Kroeger2026,
  author  = {Kröger, Wolfgang and Brettin, Leon Johann},
  title   = {Automated Vehicles: Challenging Transition from Minimal Risk
             Condition},
  journal = {IEEE Transactions on Intelligent Vehicles},
  year    = {2026},
  note    = {submitted for publication}
}

\end{lstlisting}
    \vspace{3ex}
  \end{@twocolumnfalse}
]
\setcounter{page}{0}

\maketitle

\begin{abstract}
If an automated vehicle is no longer capable of handling the dynamic driving tasks, regulations require that a minimal risk manouvre is initiated aimed at achieving minimum risk condition.
Details on measures and actions to resolve this standstill position are missing. This article provides clues for closing this gap and addresses various states in which the human intervention operator must understand the current situation and decide whether it is more acceptable to maintain the stay, e.g. on road-shoulder, or to move the vehicle back into traffic flow or ``limp it'' to another position, depending on the outcome of a health check.
Our investigations, based on a list of nearly 30 reasons and a subsequent development of ``scenario trees''  prove the complexity of the situation and the heavy difficulties faced by the operator to handle in a correct manner.
We also discuss how the technical automated driving system can be enhanced so that it stronger supports the operator or allows to even dismiss human intervention.
This deems technically feasible but limits became obvious, so that assistance from out side appears indispensable at least for the ramp-up phase of respective automated vehicle.
The non-trivial problem of transitioning the ``stranded'' vehicle needs to be addressed by further regulation for which this paper may provide useful input.

\end{abstract}


\section{Introduction}

\IEEEPARstart{A}{utomated} vehicles (AVs) offer opportunities for individual users and society as a whole.
These include better use of travel time and possible gains in comfort or possible participation in private transportation for physically or cognitively impaired persons (``inclusion''), improvement of traffic flow and traffic safety as well as technological spillover effects\cite{VDIstatus}.
According to SAE~\cite{J3016}, vehicle automation is categorized into six different levels, which differentiate between the degree of automation, the involvement of the driver and the area of application of the respective system.
With regard to the intended use, a distinction is made between privately used passenger cars, commercially used robotaxis/roboshuttles and trucks, either approved for use on public roads under specified conditions or for use in restricted private areas (so-called operational design domains, ODD). 

In terms of the state of development, significant technological progress and the first market-ready systems can be identified, with major regional differences in adaptation.
Outstanding examples include Waymo-Alphabet's robotaxi services in metropolitan areas in the west of the USA and Baidu-Apollo in China.
By 2035, only a few percent of private vehicles sold are expected to have Level 4 capabilities, fleets of robotaxis are expected to be used on a large scale in 40 to 80 cities, and up to 30 percent of new truck sales are estimated for medium-distance highway routes in the USA\cite{WorldEconomicForum}.
In this article, we limit ourselves to privately used passenger cars that are certified/allowed to drive automated on specified public roads under certain conditions according to their ODD.
The focus is on one of the safety aspects, i.e. Initialization of Minimal Risk Maneuvre (MRM) to achieve Minimal Risk Condition (MRC) and possible return of the vehicle to the traffic flow or accompanied transfer to a maintenance operation/repair (see \cref{sec: regulatory requirements} for further details).
In principle, human intervention operators or enhanced technical systems are available for control and support; their operating conditions are examined and their strengths and weaknesses/limits are explored from a technical perspective.
One notable reference here is BSI Flex 1887~\cite{BSI1887_2025}, which contains requirements for remote operators and their organizations.

To the best of our knowledge, no other related work considers the transition (``period of a system between two states'' according to \cite{Flemisch2012}) of an automated vehicle back into normal traffic from the perspective outlined in this article.
Although there are publications on the preliminary process and how the MRC is achieved by the MRM, most of these stop at the point at which the vehicle reaches the MRC.

For example, \cite{Popp2022, Ackermann2023} discuss the technical perspectives of an MRM.  
\cite{Karakaya2024} considers the transition from level 3 vehicles and stimulating ideas on how a remote operator should act in such situations.

\cite{Gyllenhammar2021} discuss the safety-related differences when executing a minimal risk manoeuvre and the mission goal cannot be reached. 
The authors distinguish between different dynamic driving task fallbacks and state that there should be a hierarchy based on the restrictiveness of the MRC. 
According to the authors of the cited paper, an MRC can be considered acceptable (or safe) if the frequency of MRM executions, the risk of the position and the resolving rate of an MRC are all taken into account.
\cite{Vu2023} expands on these concepts in the context of multiple ``collaborative'' automated vehicles, introducing the idea of ``concerted MRM'' which refers to collaborative MRM involving multiple vehicles.

Considering the potential of moving a vehicle in a limping state  \cite{Colwell2018} and \cite{Koopman2019} introduce concepts for working with vehicles with restricted capabilities based on the concept of ODDs.
In \cite{Colwell2018} ``restricted ODDs'' where the vehicle can still operate within a restricted version of the ODD are introduced.
\cite{Koopman2019} introduces ``$\mu$-ODD's''which allow for infrequent use cases with more limited capabilities, e.g. driving more slowly in freezing conditions.

A minimal risk condition does not mean, that a vehicle is in an \textit{acceptable} state after finishing the minimal risk maneuver.
This is discussed in, for example, \cite{Reschka2015, Gyllenhammar2021, Stolte2022}.
\cite{Nolte_unpublished}~following~\cite{Fleischer2023} discuss that specific words can be understood differently, but still can bring different stakeholder together and are so called ``open signifiers''.
These can for example include terms like ``safety'',``risk'' or ``risk acceptance''.
As these terms can be understood in different ways, an example of how they are perceived from a vehicle safety perspective can be found in \cite{Salem2023}.

For an automated vehicle, one major factor for acceptance is, that a vehicle is safe, but how safe the vehicle has to be is a research field on its own.
E.g., \cite{Kröger2025} proposes a quantitative measure for acceptance criteria, to complement qualitative safety requirements from a societal perspective.

Fulfilling stakeholder-dependent risk acceptance criteria, such as accepting a specific risk threshold, is a core part of safety argumentation, as stated in \cite{Loba2025}.
To address risk as an operational issue, \cite{Salem2024} proposes a framework to support the analysis of risks associated with autonomous vehicles and specify risk measures.

The article depicts - after the introduction - regulatory requirements and mandatory standards and defines key terms (\cref{sec: regulatory requirements}), introduces the neccessity of a minimal risk manoevre (MRM) to achieve minimal risk condition (MRC) and provides structured reasons for MRM (\cref{sec: reasons}) followed by dealing with the concept of remote intervention operator and related responsibilities and tasks (\cref{sec: responsibilities}). \Cref{sec: transition} focuses on transition from MRC back to normal operation or maintenance while \cref{sec: options} disputes possibilities and limits to enhance the automated technical system to support or even replace human intervention, in view of apparent situational and operational complexity and challenges, followed by discussion and conclusion (\cref{sec: discussion and conclusion}).

\section{Regulatory requirements, definitions, and mandatory standards}
\label{sec: regulatory requirements}
In recent years, a large number of laws and implementing regulations have come into force to regulate vehicle automation in order to create the framework conditions for certification and thus market launch. At EU level, for example, the Implementing Regulation \cite{EU2022_1426} and the associated Interpretation Report \cite{Ciuffo2024} should be mentioned.
In Germany and Switzerland, legislation and ordinances have been established that are based on EU requirements or are in line with them and transpose them into national law \cite{Germany_AutonomousDrivingAct2021_eng, Germany_AFGBV_2022_eng, Switzerland_OAD_2025_eng}. In Europe (as a signatory to the Vienna Convention 1968), the tried and tested concept of type approval with strict requirements is also applied to the approval of vehicles with automated driving functions\cite{EU_2018_858}. In the USA, on the other hand, the principle of voluntary self-assessment applies in the form of driverless safety reports by the manufacturer with reference to the official guidance document\cite{USDOT_NHTSA_2017_VisionForSafety}.

A network of ISO and UNECE-standards is available to demonstrate compliance with safety requirements. Some of them were created for other areas of application and then adapted to the new purpose or newly developed\cite{ISO26262, ISO_23793}. The state of the art is considered to be met if these standards, notably ISO 26262 and 2144, are fulfilled.

The various rules comprise a definition of key terms. For example, the technical system that is intended to independently manage the driving function within the defined operating range (ODD). It is called ``Automated Driving System'' (ADS) in the EU regulations, automation system (original translation: “Automatisierungssystem”) in the Swiss regulations and technical equipment (original translation “technische Ausrüstung”) in the German regulations. We use - as in other cases - the internationally common abbreviation ``ADS'' and synthesize the following definition from the slightly divergent definitions in the regulations:

\paragraph*{The ADS is a technical system that is independently capable of complying with traffic regulations, managing the dynamic driving task within its ODD, recognizing ODD limits and technical malfunctions, controlling all nominal and reasonably foreseeable (plausible) traffic scenarios within the design operational range. }

With the term ODD (Operational Design Domain) we follow the EU regulations and understand it to mean 

\paragraph*{Operating conditions under which a given ADS is specifically designed to function, including, environmental, geographical, geofenced, and time-of-day restrictions, and/ or the requisite presence of certain traffic or roadway characteristics.}

In the event that the ADS is unable to perform its above-mentioned task, reaches the limits of its ODD or is ``stuck'' the regulations require to initiate a manoeuvre aimed at minimizing risks (MRM) in traffic by bringing the vehicle to a safe stop, i.e. achieve minimal risk condition (MRC) to reduce the risk of a crash. The (fully) automated vehicle shall only leave the MRC after confirmation by self-checks of the ADS or/and the (alerted) on-board or remote intervention operator that the cause of the MRC is no longer present\cite{EU2022_1426}. 

If the remote intervention operator (\textbf{RIO}) is part of the safety concept of the manufacturer, the ADS shall provide vision systems\footnote{E.g., cameras in acc. with ISO16505:2019, ch. 6 .} of the occupant space inside the vehicle and of its surrounding to allow the RIO to assess the situation inside and outside of the vehicle. Instructions regarding the operating procedures to take over after MRM/MRC shall be described in the operating manual. 

Both safety standards which are mandatory to demonstrate compliance with the current state of technology, i.e. ISO 26262:2018 \cite{ISO26262} and 21448:2022 \cite{ISO21448}, address fallback mechanisms, MRM and transition from normal operation.
They focus on strategies and measures/requirements to mitigate risks from system malfunctions and functional insufficiency incl. \textit{foreseeable misuse}, cover driving situation till standstill.
ISO 23793, Part 2 draft document \cite{ISO_23793} covers minimum requirements and test procedures for ADS executing a MRM road shoulder stop, from initiation to termination, which aims to achieve MRC.

Detailed specifications for the transition from MRC to normal operation or maintenance are missing, in both existing regulatory requirements (EU, D, CH, US DoT) and the mandatory standards.
However, ISO 23793-1:2024 \cite{ISO_23793} states that resumption of operation after MRC shall always require human interaction with the system.

\section{Reasons for initialization of Minimal Risk Manoeuvre}
\label{sec: reasons}

The reasons for initializing and excuting a minimal risk maneuvre (MRM) by ADS are complex, multifaceted and multi-layered, but are decisive for subsequent actions.
Using external sources and our own experience, we have endeavored to collect and structure possible reasons without claiming general validity or completeness.
The causes of the reason within or outside/on the edge of the ODD serves as an essential distinguishing feature, in addition to the non-consideration of the reason in the system design (see Box 1).

\begin{figure}
\begin{tcolorbox}[colback=gray!5!white, colframe=black, ]
\textbf{I. Driving failure of the vehicle within its ODD}
\begin{enumerate}[label=\arabic*.]
    \item Loss of unrestricted driving ability as a result of:
    \begin{enumerate}[label*=\arabic*.]
        \item Accident involving the vehicle itself
        \item Hardware fault on the vehicle itself or on the sensor system 
        \item Internal software error
        \item Failure of the internal power supply
        \item Unexpected manoeuvres, violent interventions by occupants
    \end{enumerate}
    \item Interruption of the radio connection
    \begin{enumerate}[label*=\arabic*.]
        \item Long lasting
        \item Impermissible short term, including dead spots 
        \item Failure of the traffic management system including malicious cyber attacks
    \end{enumerate}
    \item Change in environmental conditions affecting driving
    \begin{enumerate}[label*=\arabic*.]
        \item Local heavy rain, snowfall, extreme \mbox{temperatures}
        \item Flooding, sudden landslides 
        \item Flash ice
        \item Local fog banks
        \item Objects on the roadway
        \item Poor markings, heavy soiling of the roadway
        \item Temporary impairments/roadworks with \mbox{special} signaling
        \item Violations of traffic rules by others 
        \item Accident situations, traffic jams
    \end{enumerate}
\end{enumerate}

\vspace{1em}
\textbf{II. Reaching/exceeding the ODD limits}
\begin{enumerate}[label=\arabic*.]
    \item Continuation of the journey due to violation of traffic regulations 
    \item Leaving the ODD without being taken over by the driver
    \item No reliable information provided by sensor system
    \item Incorrect marking or signaling
\end{enumerate}

\vspace{1em}
\textbf{III. Unexpected, unconsidered by design}
\begin{enumerate}[label=\arabic*.]
    \item Temporary external specifications 
    \item Foreseeable but unconsidered hazards 
    \item Foreseeable hazards taken into account, \mbox{parameters} exceed underlying values 
    \item Situations/combination of events beyond imagination or excluded due to extremely low probability(``epistemic uncertainties'' see \cite{Kiureghian2009, Kröger2011, Walker2003, Nolte2024})
    \item Significantly changed conditions, not taken into account via retrofits/updates
    \item Incorrect behavior of the occupants
\end{enumerate}
\end{tcolorbox}

\caption{Categorized reasons for Minimal Risk Maneuvre}
  \label{fig: failure scenarios}
\end{figure}

Both first two categories “driving failure” and “ODD border limits” are in line with current literature (e.g., ISO23793-1). We added “unexpected/unconsidered by design” reasons as another category, as (1) negligent occupant behavior is neither a direct failure nor an ODD limit and (2) unexpected trigger differ from the other 2 categories, as there is no previous  consideration reasonable or anticipation possible.

\section{Responsibilities and tasks of Remote Intervention Operator}
\label{sec: responsibilities}

Manufacturer safety concepts and regulatory requirements demand that after MRM a personalized external assistant, in this article described as Remote Intervention Operator (RIO)\footnote{This term is used in EU regulation \cite{EU2022_1426} while German official documents use the term Technische Aufsicht (technical surveillance), TA  and the Swiss ordinance uses the term Operator.}, takes over the tasks if the ADS is unable to ``bring the vehicle back to operation'' independently. The respective ISO standards leave the option of ``on-board operator'' or ``remote intervention operator'' open. 

We inherit the term RIO and assume below that an RIO is intended as a standard case who must identify the ``stranded'' vehicle and then determine whether the cause of the MRM initialization has been eliminated. If a ``health check'' confirms this, the procedure for ending the MRC can be started by returning the vehicle to the traffic flow (back to normal operation); if this is not the case, removal from the road, for example to a maintenance or repair facility, must be organized. 

The EU \cite{EU2022_1426} leaves ``instructions regarding the operating procedure to take over after MRM/MRC'' to the manufacturer.
The German and Swiss regulations \cite{Germany_AFGBV_2022_eng, Switzerland_OAD_2025_eng}, are more specific.
According to them, the RIO must check the driving maneuvres proposed by the ADS for the ``return'' to which we will restrict ourselves below, and accept them if necessary or select one of several proposed variants.
According to the German regulation, the driver must suggest an alternative maneuvre to the vehicle if this is requested by the stationary vehicle and execute it if it is accepted by the ADS, which must be deactivated if necessary.
In Switzerland, only a remote driver present on site could move the ADS away from an MRC if the ADS itself is unable to do so.

The situation faced by the RIO is highly complicated.
The demands on the RIO are challenging in terms of prompt reactions and actions to complete the MRC, which may be illustrated by the compilation, see also \cite{Dix2021}:

\begin{itemize}
    \item  Development of adequate task awareness;
    \item  Perception/understanding of the situation inside and outside the vehicle and its surroundings (situational awareness), appraisal of real-time data, communication with vehicle occupants if necessary;
    \item  Ensuring the complete elimination of the causes of MRM/MRC and/or dealing with ``partial failure'' with possible responsibility (probably in coordination with the manufacturer) for short-term continuation of the journey with reduced performance;
    \item  Development of a strategy for gradual vehicle return and re-integration into the traffic flow;
    \item  If necessary, development of an alternative driving maneuvre in accordance with safety requirements and traffic regulations, validation by ADS;
    \item  If necessary, deactivation of ADS and takeover of vehicle control; monitoring of vehicle's performance after leaving MRC;
    \item  Handing over control to the ``driver''/the reactivated ADS after solving the problem. 
\end{itemize}

For its actions, the RIO needs reliable real-time data on the status of the technical system, including the interior, and the environment, which is usually transmitted by the vehicle system.
If this not able to do this, additional sources (e.g. technical services) must be used.
In addition, the RIO requires a stable, protected radio connection to the vehicle/communication-capable vehicle occupants and to external locations, as well as acceptable latency and sufficient time reserves.

With regard to the qualification of persons who could be suited to the tasks of an RIO, the regulations provide very detailed information in some cases. These include the necessary advanced training, specific training with the manufacturer and a corresponding driving license. In addition, a basic understanding of the technical system and the problem that has arisen, the ability to pay constant attention with a monotonous stimulus frequency (vigilance), a high degree of reliability and a certain level of social and even language skills are required. However, the fields of activity and system boundaries are still quite vaguely described.

The MRC situation at the workplace of the RIO1 is characterized by the high technical and varying complexity of the overall system consisting of vehicle and automated driving function as well as the safety-related effects of its operation in dynamic traffic situations (\cite[§14]{Germany_AFGBV_2022_eng}). Depending on the concept, several vehicles may have to be monitored and/or supported at the same time, perhaps under previously untrained conditions. In addition, there is the possibility of reduced information or a lack of reliable data and its incomplete representation in the control center. The sudden demand for considerable effort after previous non- or under-demand, possibly with delayed situational awareness and impaired performance (shift work), suggests an increase in errors \cite{Bainbridge1983, Dix2021}. 

There are therefore limits to the concept of transferring the control and decision-making dilemma from a sophisticated technical system (ADS) back to a specially trained person (RIO) to compensate for its inadequacies and excessive demands. In view of the complexity of the situation and the difficulty of the task at hand, errors cannot be ruled out or are rather likely, but these may shrink with increasing experience.

\section{Transition from MRC and risk-based decision options, in general}
\label{sec: transition}

The procedure that should follow the achievement of the minimum/minimized risk condition (MRC) in detail and which is aimed at its abolition is addressed in principle in the regulations and standards, but remains largely unspecified.
This indicates a pending need for clarification (by the \mbox{manufacturer} as part of the safety concept) and a regulatory and standardization gap to be closed. 

In accordance with the EU Implementing Regulation \cite{EU2022_1426}, it is generally required that ``fully automated vehicles shall only leave the MRC after confirmation by self-checks of the ADS or/and by the on-board operator or remote intervention operator (RIO) that the cause of the MRC is no longer present''.
If this is ensured, the procedure for returning to the traffic flow (“back to normal operation'' according to ISO-standards) can be initiated, whereby ``the MRC termination takes place with the support of the ``Technische Aufsicht'' (i.e. RIO)'' \cite{Germany_AutonomousDrivingAct2021_eng}; further specifications are missing.
In the event that the reasons for MRM/MRC could not be eliminated, i.e. the defective vehicle has to be removed from the road, there is no information at all.
Equally unspecified is the handling of partial failures/reduced performance, which could consist of the loss of redundancy of the internal power supply, for example.
The EU refers to the fact that ``instructions regarding the operating procedures to take over after MRM shall be described in the operating manual'' (provided by the manufacturer).

Since the measures and procedures to leave the MRC position depend on its cause and the conditions created by it or encountered, we believe that - besides different kinds of MRC positions (see Excursus) - these must be considered when developing transition strategies with related dynamics.
Extreme road conditions that led to the MRC and must be taken into account in the return strategy may illustrate this.
In addition, there are possible errors and changed conditions during planning and implementation, so that it is advisable to think in dynamic scenarios and to develop such scenarios systematically, which to our knowledge has not yet been done so far.
To our knowledge, the current regulation and particular thoughts often end with an MRM. 

The RIO has to resolve various options based on the risk related to current MRC position and the risks involved in transitioning to another state.
Please note (see also, for example, \cite{Stolte2022} and \cite{ISO_23793}), a MRC does not necessarily result in a safe state.
It is simply the state involving the least amount of risk for the current situation.
To assess measures we use the terms “minimized risk” and “accepted risk theshold”\footnote{\cite{Stolte2022} mention this as ``Below this threshold, the risk is accepted; above the threshold, the system is considered unsafe''.}.

\begin{figure*}[ht]
  \centering
  \includegraphics[width=\textwidth]{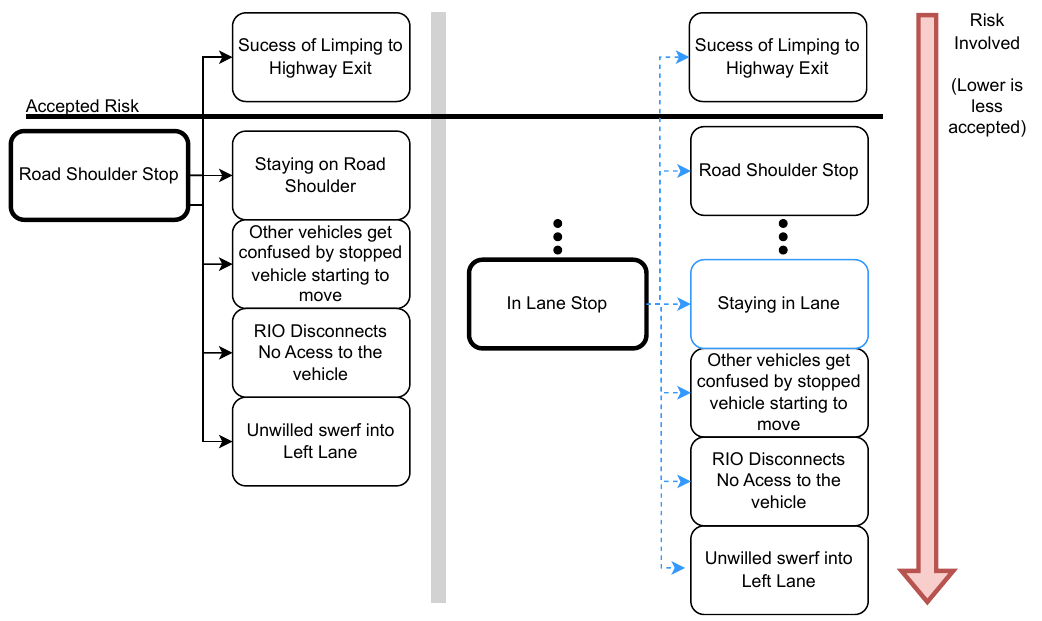}
  \caption{Risk involved in different options a RIO can take. The resulting states are for illustrative purpose and do not reflect any concrete values.}
  \label{fig: risk involved}
\end{figure*}

\begin{mdframed}[backgroundcolor=gray!15, 
  linewidth=0pt,
  frametitle={\textbf{Excursus: Transition from different MRC cases and related decision dilemma}}, 
  frametitlebackgroundcolor=gray!30, 
  frametitleaboveskip=5pt,
  frametitlebelowskip=5pt
]
We will consider \textbf{two perspectives}: the achieved end position of the vehicle and the reason for MRM.
Following ISO 23793-1, we differentiate between two categories, i.e. \textit{in-lane stops} and \textit{road shoulder} stops, distinguished between highways and urban/rural roads, the latter combined for simplicity.
The outlined main groups of reasons, namely \textit{faults}, \textit{reaching ODD border}, and unconsidered/\textit{unexpected reasons} serve as MRM-initiators which are expected to end in a minimal risk condition (MRC).
We assume that the system is able to select and execute the manoeuvre that minimizes the overall risk, but does not explicitly end in an \textit{acceptable} position regarding the risk involved.

Depending on the vehicle’s situation, we want to discuss options to further reduce risk after the MRM when the MRC is achieved.
The RIO has to decide whether the vehicle should transition back into normal traffic, stay in its position or try to “limp home”.
As all of these options can influence the risk involved we aim to draw attention to the problem by discussing some options based on the aforementioned categories, rather than offering a solution, not claiming to be exhaustive.

There is more risk involved in \textbf{stopping on a highway} than in a slower environment simply due to greater difference in speed between the stopping vehicle and normal traffic. \Cref{fig: risk involved} illustrate some decisions a RIO can take. The risk involved and the categorization of the resulting  states are for  illustrative purposes only and do not reflect any concrete values; their ordering could also be different when more meaningful risk parameters are assigned to them.

   Stopping in a \textbf{traffic lane} on a motorway is usually not considered safe. When
the system responsible for executing the MRM decides that stopping in the lane is the lowest possible risk, there should be a strong incentive to either transition the vehicle back into normal traffic or try to reach a road shoulder where it can stop more safely. 
If the vehicle stops in the lane due to a subsystem \textit{failure}, the RIO should validate whether it is possible to limp to the road shoulder.
However, various factors could increase the risk when trying to improve the situation of the vehicle:
\begin{itemize}
\item The RIO could disconnect, leaving the vehicle in an even worse situation.
\item Due to failure, the vehicle could swerve into the left lane, stopping in an even riskier position. 
\item Moving the vehicle after some time when traffic expects it to stay can also involve more risk, as it may confuse nearby traffic.
\end{itemize}

If the vehicle stops in a lane due to reaching an odd border or due to other \textit{unexpected situations}, an immediate check for trip continuation should be carried out, demanding a very high level of prioritisation. 

\vspace{\baselineskip}

Stopping on the \textbf{hard shoulder} of a road usually involves least risk, especially on a highway. However, there may be cases where stopping there is not in the accepted risk threshold (which we will not define in this article).
In this case, the RIO must decide whether to leave the vehicle in this position, which could still involve the least risk, or to transition the vehicle back into traffic or limp it to the  next motorway exit or parking zone.

This decision may vary depending on the \textbf{cause of the MRM}:
\begin{itemize}
    \item If the vehicle stops due to a failure, the question arises whether is possible to bring the vehicle back to traffic, which appears to be less risky than staying in the position or limping to the exit.  
    \item If the vehicle stops due to reaching an odd border or other unexpected reasons, the RIO has to decide whether staying there or transition back to traffic is possible. 
    \item  When the ODD border is reached, the question arises as to whether the ODD situation can be resolved. If not, the RIO can decide to limp to an exit and accept the risks involved, or stay in the current position and accept the respective risks.
    \item   If the vehicle re-enters its ODD after stopping and the RIO assumes that the same event will not recur at once (e.g. if the rain has stopped), the RIO has to decide whether transitioning back to highway traffic can involve a higher risk than in urban or rural situations. This is also applicable for the category of unexpected reasons.
\end{itemize}

So, even if the vehicle system itself performs sufficiently and a mission continuation deems possible, the RIO can decide that it is too dangerous to transition back from a lower risk to normal operation, i.e. back to high velocity traffic. The RIO may decide to let the vehicle stay on the road-shoulder, even if the situation is below the accepted risk threshold.

As in \textbf{urban or rural roads}, the velocity difference between the standing vehicle and other vehicles is not that high as in a highway setup, making the overall stopping situation less critical. Further,  a vehicle in stopping lane or aside should be more common, which may result in the potential for adequate attention for stopped cars of other traffic participants.
 
If the vehicle executes a MRM in a \textbf{traffic lane}, the RIO should check if it is possible to move the vehicle to the hard shoulder. However, the potential for an accident, resp. the risk involved with staying in lane is lower than in the highway setting. Due to safety reasons, the RIO should aim to move the vehicle away from the traffic anyway. If the failure is such that moving the vehicle would increase the risk, it should stay in the lane. If the vehicle stopped in lane due to ODD border reached, or due to other unexpected reasons, the RIO must decide whether it is possible to slowly maneuver the vehicle to the hard shoulder, or if it this too risky to move the vehicle out of its ODD. 
  
 When a vehicle stops on the \textbf{hard shoulder}, the risk of staying in that position is slightly above the accepted threshold:
 the RIO can allow the vehicle to remain in this position without having to decide immediately what to do instead.
 If it is safe to return the vehicle to normal traffic and the reasons for the MRM have been resolved, this could be an option for resuming normal operation.
 If not, there should be enough time to arrange for an operator to move the vehicle, e.g., to a garage for repair or maintenance. 
\end{mdframed}

\section{Options for actions to terminate specified Minimal Risk Condition: Scenario trees }
\label{sec: options}

To further detail actions, potential scenarios where analysed after completion of the MRM and the vehicle has entered a minimal risk condition, aimed at leaving this position.
The states in which the vehicle can be found and actions available to the vehicle and the Remote Intervention Operator (RIO) were taken into account.

\begin{figure}[ht]
    \centering
    \includegraphics[width=\linewidth]{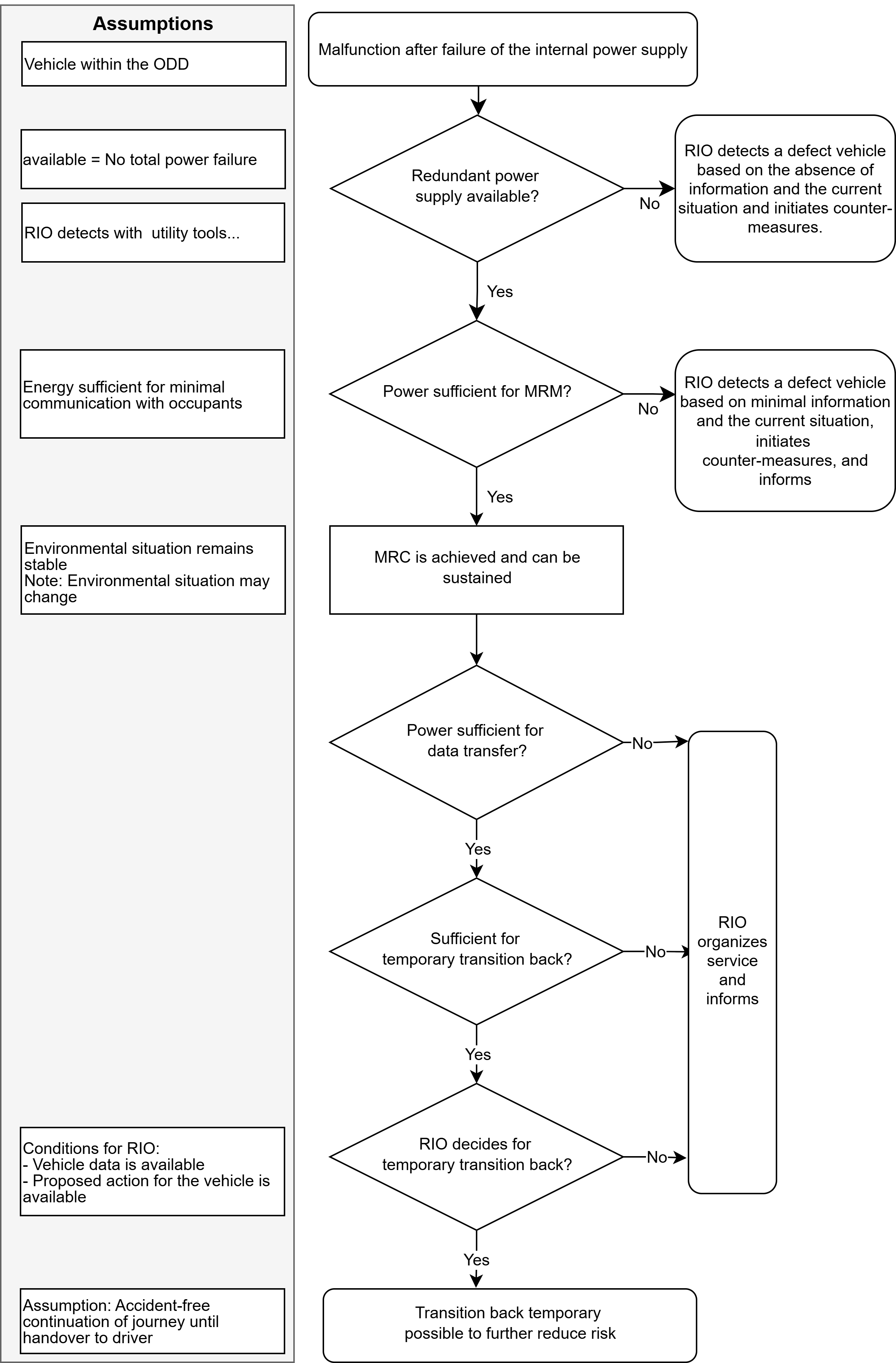}
    \label{fig: scenario tree}
    \caption{The scenario tree shows examples of decisions that a RIO has to make.}
\end{figure}

We used the inductive flow chart method as an example for a total of five of the many MRM/MRC reasons (Box 1) and created so-called scenario trees in two brainstorming sessions with a maximum of 8 knowledgable participants. \cite{SATW}
This widely used technique starts from the occurrence of a defined cause, successively assumes the existence or non-existence of conditions in the form of unweighted ``binary questions'' (yes/no, can be/cannot be) and ultimately leads to different end states; assumptions relevant to the result were necessary and made.
Such a ``scenario tree'' as a result is shown as an example for the MRM/MRC reason ``Failure of the internal power supply'' in \Cref{fig: scenario tree}.

The following findings were derived from the development of several scenario trees:

\begin{itemize}
  \item The analytical method used on a trial basis proved to be suitable and efficient.
  \item The results are specific, but allow generic conclusions to be drawn.
  \item Assumptions were necessary in order to be able to generate a manageable number of plausible scenarios; they must be checked for resilience, their large number demonstrates the situational and cognitive complexity.
  \item The cases investigated, triggered by technical failures (e.g. loss of the regular power supply or data transmission), demonstrate the importance of the redundancy of sensitive systems or high latency to compensate for a lack of redundancy.
  \item RIO must be technically supported to be enabled to recognize the failure of AV/ADS  within a reasonable period of time and to prepare actions mentally.
  \item In most of the cases investigated, MRM can be safely initiated by AV/ADS and MRC can be achieved.
  \item However, there were also cases in which the vehicle would have continued to drive under conditions that did not conform to the traffic rules/design, which, inter alia, raises  the question of the need for continuous external monitoring.
  \item Avoiding malicious or accidental operating actions is a high priority for AV/ADS, so that accidental contact with driver's controls does not lead to an unintentional cessation of automated mode.
  \item Many scenario chains end with temporary return, sometimes with partially defective ADS, to reduce risk, which places an additional burden/responsibility for the RIOs (see below).
  \item MRC (e.g., road-shoulder stop) may not be permanently safe as a result of a dynamic change in the traffic situation and should be ended as quickly as possible (see e.g., \cite{Stolte2022}).
  \item The scenario trees also provide indications of efficient system improvements (``smart'' door locking systems as an example).
\end{itemize}

 The large number of cases identified and possible sequences initiated let us conclude that some  MRMs are to be expected in the ramp-up phase of automated vehicles, which seems to make technical supervision/remote intervention and also on-site interventions indispensable.
 Existing control centers of fleet operators (such Waymo Robotaxis) or system providers/manufacturers (such as Aurora) with accumulated experience are of undeniable value in terms of ``lessons-learned'' and ``best practice'', if shared.

Although required by current regulation and being part of the safety concept of manufactures in a general way, a MRM might not be the solution for every abnormal event.
Moreover the current way of thinking process often stops at the solution of executing a MRM whereas a decision is pending afterwards, if the vehicle is in a situation which cannot be considered as safe.
 
By example, assuming that the MRM system works as designed for and the vehicle stops in an appropriate MRC it could still be below the accepted risk threshold.
If the RIO has no option to resolve this state, it should be discussed if such an MRM should be executed in the first place. 

Probably the AV's ODD should be adapted, so that a MRM is only possible when an acceptable risk level can be ensured, saying that even when the ADS is capable of driving in a certain area the vehicle should not do so, if the MRM would end in a non acceptable state.
Another option could be to transition the vehicle into a different limping state or move the vehicle a little further and stop in a more acceptable state although the duration can be rather long compared to one manoeuvre, which should be executed urgently.

Thus, there may be cases where the RIO must choose between two unacceptable options: remaining in an unacceptable position or moving the vehicle with an unacceptable risk of causing harm by the maneuver.
This difficult decision could incorporate legal problems.

\section{Enhanced automated driving system}
\label{sec: enhanced automated driving system}

The concept of the personalized take-over (RIO) of a control and action function to compensate for inadequacies and imponderabilities of the automated technical system (ADS) seems paradoxical, as it was created to eliminate humans as the main cause of accidents.
As explained, it is also associated with considerable difficulties and challenges and, given the complexity of the situation and task, has limited prospects of success (see, e.g., \cite{Bainbridge1983} for similar problems).
Accordingly, the question arises as to whether a technical system, the ADS, can be enhanced to such an extent that human intervention from outside can be more significantly supported or even dispensed with.
This subsection aims to explore ideas on how this could be achieved.

In any case, manufacturers, providers, and researchers are striving for the highest possible reliability and functional safety of the ADS and its continuous improvement.
The number of MRM/MRC and necessary personalized interventions (by RIO), including actions for transition from MRC, should be further reduced, concomitantly damage to the image and loss of confidence in the technology should be avoided, even the visionary goal of zero risk appears at a high level of abstraction aimed at eliminating traffic fatalities at all \cite{Blumenthal2020}.
A proposal has been made to supplement qualitative safety requirements such as to avoid unreasonable risk by aggregated quantitative target values \cite{Kröger2025}.

We assume that generally valid core principles and associated measures of reliability theory and technology are followed and consistently implemented in these endeavors, such as:

\begin{itemize}
  \item Continuous self-monitoring with reporting of anomalies, use of fail-safe and fail- operational mechanisms and graceful degradation behavior~\cite{Stolte2022}).
  \item Integrated redundant design of essential hardware (sensors, power supply) and software systems (different, mutually checking algorithms);
  \item Hazard and reliability analysis deploying well-established formal methods, initially using generic, later specific data; 
  \item Rigorous testing (bench, virtual, geofenced, real world), simulation of traffic and accident situations; compliance with best practice.
\end{itemize}

In addition, AI-based learning, continuous software updates and the comparison of safety assessments with real-world information before commissioning.

An even more rigorous use of these principles and measures, proven to be possible and economically justifiable, would serve the above-mentioned goals aimed at continuous improvements.
Specifically, in order to reduce the probability of malfunctions of the vehicle/ADS within its ODD as a result of hardware and software failures, strengthen its end-to-end redundancy, including the power supply.
Where appropriate, diversity should be sought to avoid common cause failures (CCF)
\footnote{Experience from other domains, notably the nuclear sector~\cite{Dona2022}, has shown that extremely low failure probabilities of vital safety systems cannot be achieved only by increasing the number of components/trains a but call for their diversity, e.g., different designs of diesels of a redundant emergency power system.}.
In addition, an optimized structure of the entire perception system consisting of sensors with different ranges and achievable accuracy (camera, lidar, radar) and fusion of generated data should be aimed at.
This could also contribute to greater robustness against serious changes in environmental/driving conditions and would allow a possible extension of the ODD including the avoidance of respective MRM/MRC reasons.
An extension of the simulation to very rare situations and events and the use of advanced methods, allowing to better represent the complexity of the system\footnote{Methods following a “systems approach” such as Complex Network Theory (CNT), agent-based modeling (ABM), Systems Theoretic Process Analysis (STPA), see also \cite[chapter 4]{Kröger2011}.}, could help to reduce the degree of ``unexpected'', notably by analyzing and including ``edge scenarios'' in the system design.
The protection of communication channels and data pipelines against malicious intrusion should be strengthened where appropriate and possible.
Particular attention should be paid to the adequate handling of the human-machine interface.

In addition to reducing the number of MRMs initiated and MRCs reached, an enhanced technical system with an extended range of tasks and  further increased reliability could mitigate or even eliminate some of the tasks that the RIO is currently expected to perform.
E.g., an ADS whose function has been adapted with an associated ``optimized hardware and software (algorithms)'' perception system could carry out the vehicle's ``self-health check'' and largely develop the strategy for leaving the MRC-position itself and carry out the necessary measures.

Despite all the confidence in technical possibilities, risks may be involved.
Limits are also recognizable here, they are of a fundamental technical-physical and also of monetary nature.
Further increase in complexity needs to be expected, understood and mastered, uncertainties remain but could be reduced thanks to growing empirical data and its systematic evaluation as well as learning of a different kind.
However, residual uncertainties are probably unavoidable.
In addition, acceptance by vehicle users appears to be open to question.

\section{Discussion and Conclusion}
\label{sec: discussion and conclusion}

Automated vehicles (AV) have a technical system (ADS) that is capable of handling the dynamic driving task.
This includes ``plausible'' traffic scenarios within the area of use and under conditions for which it is designed and allowed to operate (operational design domain, ODD), in accordance with the traffic rules, and of recognizing the limits of its ODD and technical faults.
If the ADS is unable to perform its task and reaches the limits of its ODD, the regulations require that it can initiate a risk minimization manoeuvre (MRM) and places the vehicle in minimal/minimized risk condition (MRC).

In this case, manufacturers' safety concepts and regulatory requirements demand that an external qualified person, \mbox{variously} referred to as a remote intervention operator (RIO), provides support and takes over required tasks.
The RIO must check the ``stranded'' vehicle for the cause of the MRM and then ensure MRC’s termination, I.e. depending on the outcome of the check, return the vehicle to the traffic flow or organize its removal to the maintenance (repair) operation.

This concept for compensating for the inadequacies of the ADS and remaining uncertainties by ``humans'' seems paradoxical, as the technical systems was created aiming to eliminate the “human factor” as the dominant cause of accidents \cite{Dix2021} following \cite{Bainbridge1983}. 

Existing regulations entail rather general descriptions of the situation at the RIO’s workplace and the actions to be carried out, in contrast to some detailed requirements regarding operator qualification and suitability.
The reasons for MRM and possible MRC states are only mentioned in general terms, details on measures to resolve the MRC position are completely missing, which indicates a shortcoming of current thoughts and a regulatory deficit.
This paper provides clues for closing this gap. 

A structured list of almost 30 reasons was compiled that lead to the triggering of MRM which are important to know for subsequent measures and actions.
Our ``scenario tree'' approach is an example of how event-dependent and varied those are in terms of their requirements and timing.
The technique used on a trial basis has proven to be suitable and efficient without making other approaches obsolete.
Some of the specific results obtained can be generalized and used for system improvements.
 
Our attempt to characterize the conditions at the RIO's workplace has highlighted the complexity of the situation and the overall system, the difficulty of the task and the challenges faced by the operator.
Limits to the implementation of the concept became apparent, especially as the ADS (including the sensor system) of the vehicle serves as the primary data source when stationary.
 
We investigated the question of whether a technical system/ADS could be upgraded/enhanced in such a way that external human intervention is significantly stronger supported or even no longer necessary.
It seems that extending the range of tasks and further increasing reliability in a targeted manner is technically very challenging, but possible.
However, limits must be taken seriously and - in addition to cost increases - considerable residual uncertainties are unavoidable, although these could be successively reduced by exploiting growing operating experience.

The long-term goal of ongoing ADS development (with a view to privately used SAE L 4, potentially L5 vehicles) should be to be able to dispense with personalized assistance from outside even when the automated vehicle has to transit from MRC position.
Until this is the case, it will probably be necessary, may be required by law or desired by the operator, especially as some MRMs are to be expected in the ramp-up phase of respective AV while on-site interventions also appear to be indispensable.

Certainly, regulatory gaps still need to be closed, i.e., requirements and conditions for necessary actions need to be specified and cause-dependent action scenarios need to be developed.
This paper aims to provide some points of interest for this.
However, our thoughts are limited to privately used ego vehicles and with regard to fleet operation additional challenges appear and considerations are necessary, e.g., space could be unacceptable for road shoulder stops of multiple vehicles.  

\section*{Acknowledgement}
We would like to thank the members of the \textit{SATW Topical Platform group} for their contribution: B. Gerster (once FH Bern), T. Beutler (Helbling Technik),  M. Deublein (BFU), C. Holzner (SATW), S. Huonder (ASTRA), B. Lindner (Planzer), J. Michel (once PostAuto), T. Probst (U Fribourg), T. Rücker (auto-schweiz), R. Schneider (once SwissRe).

\renewcommand*{\bibfont}{\footnotesize} 

\printbibliography

\begin{IEEEbiography}[{\includegraphics[width=1in,height=1.25in,clip,keepaspectratio]{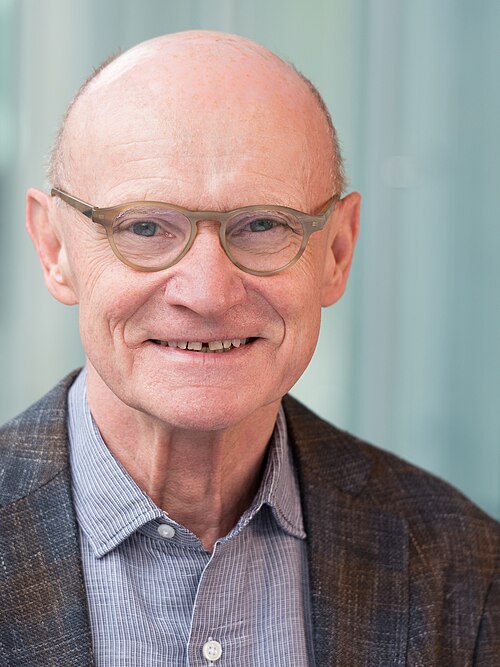}}]{Wolfgang Kröger} received the Diploma in \mbox{mechanical} engineering and Ph.D. degree in nuclear safety from the RWTH Aachen, Germany. In 1990, he became full professor for safety technology at the ETH Zurich and focused his seminal work on risk assessment of complex technical systems. In parallel, he served as Founding Rector of the International Risk Governance Council (IRGC). He was executive director of the ETH Risk Center, after his retirement in 2011. Currently, he works as scientific advisor of prestigious institutes and organisations. He is a member of the Swiss Academy of Engineering Sciences (SATW) and leads its topical platform "Autonomous Mobility" and is a member of the project on future energy systems (ESYS) of three German academies. He has been awarded Distinguishesd Affiliated Professor and fellow by TU Munich. He has published a multitude of scientific papers, books, and edited volumes dedicated to the vulnerability/resilience of interdependent vital infra-structure systems and, most recently, to safety/security related issues of automated driving systems.
\end{IEEEbiography}

\begin{IEEEbiography}[{\includegraphics[width=1in,height=1.25in,clip,keepaspectratio]{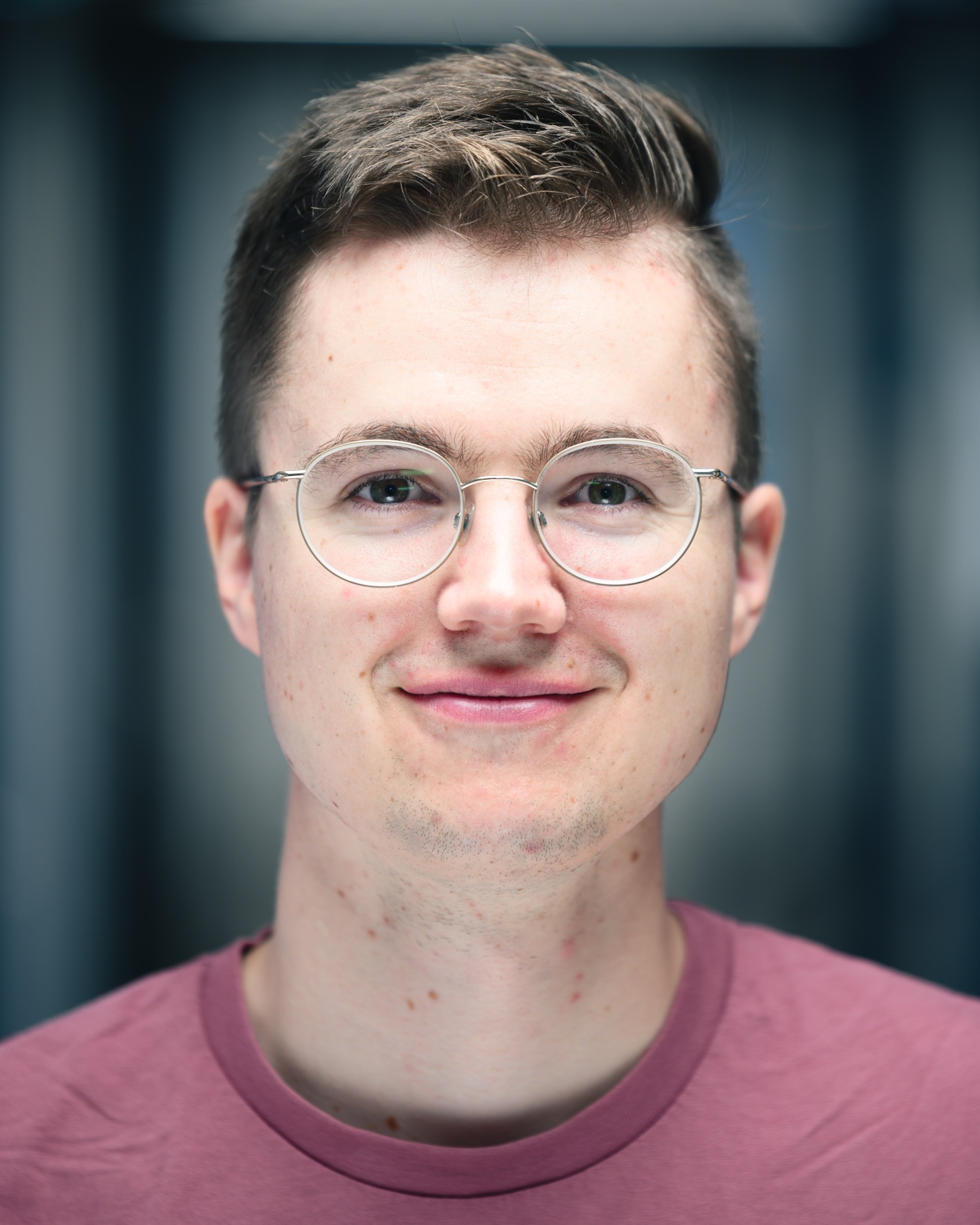}}]{Leon Johann Brettin} received the B.Sc. degree in computer science from Ostfalia \mbox{University} of Applied Sciences, Wolfenbüttel, Germany (2017) and the M.Sc. degree in computer science from \mbox{Technische} Universität Braunschweig, Germany (2021).
He is currently a Research Associate with the Institute of Control Engineering, Technische Universität Braunschweig. His main research \mbox{interests} include the areas of teleoperated vehicles and human-machine-interfaces.
\end{IEEEbiography}

\end{document}